\documentclass[11pt,a4paper]{article}

\usepackage[utf8]{inputenc}
\usepackage[T1]{fontenc}
\usepackage{mathpazo}
\usepackage{microtype}
\usepackage[margin=1in]{geometry}
\usepackage{amsmath,amssymb}
\usepackage{booktabs}
\usepackage{array}
\usepackage{enumitem}
\usepackage{setspace}
\usepackage[font=small,labelfont=bf]{caption}
\usepackage{hyperref}
\title{Neutrinos as Quantum Information Carriers?\\
A Gap Analysis Against the DiVincenzo Criteria}
\author{Ramanuj Awasthi\\
\small Chhatrapati Shahu Ji Maharaj University\\
\small \href{mailto:ramanujawasthi19@gmail.com}{\texttt{ramanujawasthi19@gmail.com}}}
\date{October 25, 2026}

\begin{document}
\maketitle

\begin{abstract}
Photons, trapped ions, neutral atoms, and superconducting circuits are all well-known quantum information platforms. Neutrinos offer a unique possibility, displaying quantum interference (flavour oscillation) over thousands of kilometres, yet interacting so weakly that they can travel through the Earth with minimal interaction. This short paper is a gap analysis. We use information from the existing literature to compare neutrinos to the seven DiVincenzo requirements. Firstly, there are five requirements to do with computation and secondly two requirements to do with communication. We analyse where the gaps are for each requirement. Overall, we find that neutrinos have high fidelity and good coherence; however, they do not have a clear qubit definition and initialisation. They do not have universal gates, individual qubit readout, or stationary-to-flying conversion, meaning they either fail or are very poor. Therefore, we think they are best considered as an exceptional flying information carrier; however, they are very difficult to control. We have outlined all of the gaps which can be used to inform future work.
\end{abstract}

\noindent\textbf{Keywords:} neutrino oscillations, DiVincenzo criteria, flying qubit, quantum communication

\section{Introduction}

A qubit is not an object but a role that any quantum system with two well-controlled, measurable, and coherent states can perform. In practice, this can be done using various physical systems such as photons (polarisation or path), trapped ions, neutral atoms (internal states), and superconducting circuits, but each has pros and cons.

Neutrinos are quantum particles that have flavour states (electron, muon, tau) which are superpositions of mass states. They also have some of the longest-range quantum interference effects seen in nature due to flavour oscillation. However, they are extremely difficult to control and measure due to their weak interaction. Here we ask a simple question: if photons can be used to carry quantum information, what would need to be true for neutrinos to do the same? Where does this idea break down?

Using DiVincenzo's criteria~\cite{divincenzo2000}, we go over each criterion and explain what evidence we have, whether we pass or not, and identify the main issue. We aim to understand this problem to allow future work to target these issues. We do not demonstrate that neutrinos can be used as working qubits.

\section{The DiVincenzo criteria}

The five computation criteria are: (1) a scalable system with well-characterised qubits; (2) the ability to initialise qubits to a simple state; (3) coherence times much longer than gate times; (4) a universal set of quantum gates; (5) a qubit-specific measurement. The two communication criteria are: (6) the ability to interconvert stationary and flying qubits; (7) the ability to transmit flying qubits faithfully between locations~\cite{divincenzo2000}. Since the neutrino is naturally a flying particle, criteria 6 and 7 are especially relevant.

\section{Background}

\noindent\textbf{Neutrino Oscillations:} A particle containing a specific flavour composition can be written as a superposition state of 3 mass eigenstates defined by the PMNS mixing matrix. As the neutrinos propagate, each mass state picks up an additional phase, causing the flavour composition to change. This was first strongly demonstrated by using atmospheric neutrino data in Super-Kamiokande~\cite{superk1998}. Current global data fits well under the three-neutrino scheme~\cite{pdg}.

\medskip
\noindent\textbf{Quantum Information:} Neutrino flavour oscillations can also be described as flavour entanglement between the single particle~\cite{blasone2009}. There has been work on using quantum computers to simulate neutrino systems using qubits and qutrits~\cite{turro2024,spagnoli2025}.

\medskip
\noindent\textbf{Neutrino Communication at Fermilab:} The MINERvA collaboration successfully demonstrated a neutrino communication link in 2012. The NuMI neutrino beam was pulsed to create a binary message which was received by the MINERvA detector 1.035~km away. The neutrinos travelled through approximately 240~m of solid rock. Data was decoded at roughly 0.1~bit/s with a bit error rate of approximately 1\%. This showed that neutrinos can be used to transmit classical information through a medium that would block all other carriers. However, this required a large particle accelerator and detector weighing several tonnes. The communication rate was many orders of magnitude slower than optical links and no quantum aspects of the communication channel were investigated~\cite{stancil2012}.

\section{Criterion-by-criterion gap analysis}

\subsection{Criterion 1: qubits (partial)}
The nature of the neutrino is a qutrit due to it being in 3-dimensional flavour space. However, when there is a large dominant mass split (as is the case in atmospheric 2-flavour fits~\cite{superk1998}), this can be approximated to a 2-level system. Unfortunately, as the baseline increases, a third flavour will appear. There is no known way to create a scalable, interactive register of neutrinos. Neutrinos have very little interaction with each other.

\subsection{Criterion 2: initialisation (partial)}
The neutrinos produced will have a specific composition of flavours based on the source. This provides ensemble initialisation rather than individual state preparation. This was why many neutrinos were sent in bursts at Fermilab~\cite{stancil2012}. Unfortunately, there is no individual state preparation.

\subsection{Criterion 3: coherence (promising)}
Neutrinos have been observed oscillating for long distances ranging from hundreds of km to approximately 13,000~km across the Earth~\cite{superk1998}. A study of MINOS data showed the violation of the Leggett--Garg inequality over 735~km, reported as the longest distance such a test has been performed~\cite{formaggio2016}. However, using wave-packet models, the oscillations will be suppressed after a certain point due to the separation of the mass-state wave packets~\cite{akhmedov2022}. The coherence observed is only that of an ensemble detected and not that of an individual qubit stored. There are no gates available to define a ratio between the coherence and gate time.

\subsection{Criterion 4: universal gates (fail)}
To have universal gates in a system, you must have control over single-qubit rotations and the ability to entangle. The only known evolution for neutrinos is that which occurs naturally (vacuum oscillations and modification in matter). Unlike other platforms, you cannot programme the neutrinos using laser/microwave/optical components. We are not aware of any neutrino gates available that can be programmed. Due to the neutrinos going through the weak interaction, it is extremely difficult to have control.

\subsection{Criterion 5: measurement (very poor)}
The flavours can be identified through charged current interactions which lead to the production of the corresponding charged lepton~\cite{superk1998}. However, this is a destructive measurement and only those neutrinos which interact will be measured. The efficiency of the measurement is extremely low, even with using a multi-ton detector only a small number of neutrinos were detected at Fermilab~\cite{stancil2012}. An order-of-magnitude calculation can be done to find the probability of the neutrino interacting, $P \approx \sigma N_A \rho\, d$ (where $\sigma$ is the cross section, $N_A$ is Avogadro's number, $\rho$ is density, and $d$ is distance): around $10^{-11}$ to $10^{-8}$ for neutrinos of about 1--100~GeV in a detector tens of metres to 1~km thick (assuming a cross section that grows linearly with energy). The readout efficiency would need to increase by many orders of magnitude.

\subsection{Criterion 6: stationary--flying conversion (fail)}
There is no known way to store the state of the neutrino or transfer between a stationary qubit and a neutrino. Unlike photons in cavities and fibres, neutrinos cannot be trapped. Without this interface, the neutrino can only be used as a one-way communication device.

\subsection{Criterion 7: faithful transmission (strong)}
The neutrinos excel in this area. They travel through the Earth to be detected (atmospheric upward-going neutrinos)~\cite{superk1998} and have been used to transmit data through rock at Fermilab~\cite{stancil2012}. However, it is unknown if quantum information can be transferred rather than classical bits.

\subsection{Summary}

\begin{table}[h]
\centering
\renewcommand{\arraystretch}{1.25}
\begin{tabular}{@{}p{0.34\textwidth}p{0.15\textwidth}p{0.40\textwidth}@{}}
\toprule
\textbf{Criterion} & \textbf{Verdict} & \textbf{Main gap}\\
\midrule
1. Scalable qubits & Partial & No multi-neutrino register\\
2. Initialisation & Partial & Ensemble, not per particle\\
3. Coherence & Promising & Single-qubit coherence untested\\
4. Universal gates & Fail & No programmable control\\
5. Measurement & Very poor & Destructive, very low efficiency\\
6. Stationary--flying conversion & Fail & No storage or interface\\
7. Faithful transmission & Strong & Quantum link not demonstrated\\
\bottomrule
\end{tabular}
\caption{Qualitative summary. Verdicts are the author's judgements based on the cited literature, not a quantitative score.}
\end{table}

\section{Discussion and possible directions}

The neutrino scores best where photons are also strong (transmission, coherence) and worst where controllable hardware is required (gates, readout, memory). The same weak interaction explains both, so the neutrino is more naturally a specialised flying carrier than a computing qubit. Possible directions that follow from the gaps are:
\begin{itemize}[itemsep=2pt]
\item improving classical neutrino links (richer encoding, better detectors, error-correcting codes) for settings where other carriers fail;
\item theory work on matter effects, decoherence and three-flavour Leggett--Garg or entanglement-witness tests;
\item a systematic search for any proposal of neutrino gates, storage or conversion;
\item studying whether a Bell-type or other quantum test is possible on a neutrino link.
\end{itemize}
What would change the verdict is (i) a controllable interaction giving programmable gates, (ii) readout efficiency many orders of magnitude higher, and (iii) any storage or conversion interface. None is currently known.

\section{Limitations}

This is a perspective based on literature, with no new experimental data. The verdicts are qualitative, and the absence of programmable gates and storage methods rests on not finding any such work in our survey; it should be rechecked with a systematic search. Parameter values and the readout estimate are order-of-magnitude only.

\section{Conclusion}

Against the seven DiVincenzo criteria, the neutrino is an excellent flying information carrier but a very poor computing qubit. Its coherence and Earth-crossing transmission are remarkable, while its weak interaction blocks gates, efficient readout, and memory. We offer this gap map as a starting point for later, more detailed work.

\end{document}